\documentstyle{article}
\date{}

\begin{document}
\title{An Imrprovement on RPA Based on a Boson description}
\author{{Shi-Lin Zhu\    \ Zhi-ning Zhou\    \ Ze-sen Yang}\\
  {Physics Department, Peking University, Beijing, 100871, P.R.China}}

\maketitle

\begin{center}
\begin{minipage}{120mm}
\vskip 0.8in
\begin{center}{\bf Abstract}\end{center}
{We use a solvable model to perform modified Dyson mapping and reveal the
unphysical-state effects in the original Random Phase Approximation (RPA). We
then propose a  method to introduce the RPA and improve it based on the Boson
description.}
\end{minipage}
\end{center}

\section *{I.Introduction}
\indent
\par
    The RPA is often used to describe collective excitations. In nuclear
physics it is widely employed to define surface oscillation and pair
oscillation\cite{s1}. Recently the RPA has also been used to study mesons in
terms of the quark-antiquark degrees of freedom\cite{s2,s3}.
\par
    The RPA builds bosonic excitations with fermions so it includes
unphysical state effects. In view of its important role in practice, it is
helpful to explore convenient approaches for improvement. In this paper we
will  develop such a method
by studying a simple solvable model with the help of the modified Dyson
mapping\cite{s4,s5}. We will introduce the RPA in a different way and explain
it as the zero order approximation of a general Boson description. Then only
very low orders of
the correction terms will be included so as to preserve the simplicity of the
method. It will be seen that  our method are both quite powerful. Although
the model
is very simple our method can be, in principle, generalized approximately.

\section *{II. Model and Boson Description for Particle-Antiparticle Pair
                States}
\indent
\par
    Consider a simple system of identical Fermions and their anti-particles
and suppose that each particle has only two independent states. The
Hamiltonian is assumed to be:
$$
H=\epsilon\sum\limits_{\mu=\pm1}(b_{\mu}^{\dag}b_{\mu}+d_{\mu}^{\dag}d_{\mu})
+R(b^{\dag}_1 d_{-1}^{\dag}b_{-1}^{\dag} d_1^{\dag}+d_1 b_{-1}d_{-1}b_1) \, ,
$$
where $b_{\mu}^{\dag}|0\rangle$ and $d_{\mu}^{\dag}|0\rangle$ stand for the
particle states and anti-particle states respectively. The state vector of
the ground state of the system is
$$ |\Psi_0\rangle = x_1 |0\rangle
+ x_2 b^{\dag}_1 d_{-1}^{\dag} b_{-1}^{\dag} d_1^{\dag} |0\rangle \, ,$$
with
$$\frac{x_2}{x_1}=\frac{E_0}{R}=\frac{2\epsilon}{R}-\sqrt{
(\frac{2\epsilon}{R})^2+1} \, ,$$
where $E_0$ is the ground state energy:
$$E_0=2\epsilon-\sqrt{(2\epsilon)^2+R^2} \, .$$
The particle-antiparticle pair states $b_1^{\dag} d_{-1}^{\dag} |0\rangle$
and $b_{-1}^{\dag} d_1^{\dag} |0\rangle$ are also the exact eigen states of
$H$ belonging to the eigenvalue $2\epsilon$. Namely,
$$\left \{
\begin{array}{l}
H b_1^{\dag} d_{-1}^{\dag} |0\rangle
= 2\epsilon b_1^{\dag} d_{-1}^{\dag} |0\rangle \, ,\\
H b_{-1}^{\dag} d_1^{\dag} |0\rangle
= 2\epsilon b_{-1}^{\dag} d_1^{\dag} |0\rangle \, .
\end{array}
\right.   $$
\par
    On the other hand the RPA excitation of a particle-antiparticle pair
takes the following form
$$ Q^{\dag}_1=X b_1^{\dag} d_{-1}^{\dag} + Y d_1 b_{-1} \, , $$
$$ Q^{\dag}_{-1}=X d_1^{\dag} b_{-1}^{\dag} + Y b_1 d_{-1} \, , $$
where X,Y are real numbers and to be found from the RPA equation:
$$\left \{
\begin{array}{l}
[H,Q^{\dag}_{\mu}]\approx \omega_{RPA}Q^{\dag}_{\mu}\\
\left [Q_{\mu},Q^{\dag}_{\nu}\right ] \approx \delta_{\mu\nu}
\end{array}
\right. $$
The solution is :
$$\left \{ \begin{array}{l}
\omega_{RAP}=\sqrt{(2\epsilon)^2 -R^2}\\
\frac{Y}{X}=\frac{2\epsilon}{R}-\sqrt{(\frac{2\epsilon}{R})^2-1}
\end{array}  ,(X^2-Y^2=1) \right. \, .$$
$\omega_{RAP}$ is the excited energy with respect to the RPA ground state
which is determined by the condition:
$$ Q_{\mu}|\Psi^{RPA}_0\rangle =0 \, .$$
One gets from this
$$ |\Psi^{RPA}_0\rangle = X_1 |0\rangle
+ X_2 b^{\dag}_1 d_{-1}^{\dag} b_{-1}^{\dag} d_1^{\dag} |0\rangle \, ,$$
with
$$\frac{X_2}{X_1}=-\frac{Y}{X}=-\frac{2\epsilon}{R}+\sqrt{
(\frac{2\epsilon}{R})^2-1} \   \ (X_1^2+X_2^2=1) \, .$$
It can be seen clearly that the RPA  worsens when
$\frac{2\epsilon}{R} \rightarrow 1$ and breaks down when
$\frac{2\epsilon}{R} <1$.
\par
    Since the RPA operators $Q_{\mu} \, ,Q^{\dag}_{\mu}$ are assigned to satisfy the
standard Boson commutation rule, it should be convenient to analyse the RPA
within an exact Boson description. For this purpose, it is enough to consider
the Boson image of the invariant subspace $V_F$ of $H$, which is spanned by
the states:
 $$ |0\rangle \, ,
    b_1^{\dag}d_{-1}^{\dag}|0\rangle \, ,
    b_{_1}^{\dag}d_1^{\dag}|0\rangle \, ,
    b_1^{\dag}d_{-1}^{\dag}b_{_1}^{\dag}d_1^{\dag}|0\rangle \, . $$
We now employ the modified Dyson mapping to set up a 1-1 correspondence
between the fermion state subspace $V_F$ and a Boson state subspace $ V_B$.
The latter is spanned by the four Boson states
    $$|0) \, ,\ \ B_1^{\dag}|0) \, ,\ \ B_{-1}^{\dag}|0) \, ,
    \ \ B_1^{\dag} B_{-1}^{\dag}|0) \, ,$$
where $B_{\mu}^{\dag}$ are Boson creation operators, and
$B_{\mu}$ are the corresponding annihilation operators. They of course
satisfy the standard Boson commutation rule.
 These four Boson states
are to be described as the Boson images of the above four Fermion states.
Therefore, $|0)$ has the same quantum numbers as $|0\rangle$ and satisfies
          $$B_{\mu}|0)=0 \, ,\ \ \ \ (0|0) = 1 \, ,$$
and $B_1^{\dag}|0) $ and $B_{-1}^{\dag}|0)$  carry the quantum numbers of
$b_1^{\dag}d_{-1}^{\dag}|0\rangle$ and $b_{-1}^{\dag}d_1^{\dag}
|0\rangle$ respectively. The correspondence can be expressed as
          $$ \Phi  =  \Gamma_B U \Psi \, , $$
          $$  \Psi =  U^{\dag} \Gamma_B\Phi  \, , $$
where
$$U=\langle 0|e^{B_1^{\dag} d_{-1} b_1 + B_{-1}^{\dag} d_1 b_{-1} } |0)\, , $$
$$U^{\dag}
 = (0|e^{b_1^{\dag} d_{-1}^{\dag} B_1 + b_{-1}^{\dag} d_1^{\dag} B_{-1}}
   |0\rangle \, . $$
$\Gamma_B$ is the projection operator to the subspace $V_B$. Namely:
$$\Gamma_B\Phi=\left \{
\begin{array}{l}
\Phi \        \ \mbox{when  }   \Phi \in V_B \\
0  \    \  \mbox{when  }    \Phi \perp V_B .
\end{array} \right. $$
According to this correspondence, one can easily find the Boson image of the
Hamiltonian and write it as
$$ H_B=\Gamma_B\{2\epsilon\sum\limits_{\mu}B_{\mu}^{\dag}B_{\mu}+
R(B_1^{\dag}B_{-1}^{\dag}+B_{-1}B_1)\}\Gamma_B \, . $$
This Boson Hamiltonian and  the $V_B$ construct an exact Boson description
for the particle-antiparticle pair states. Each Boson state of the subspace
$V_B$ is a physical one, and a Boson state with components outside $V_B$ is
partially or totally unphysical. In this description, the 1-Boson eigenstates
of $H_B$ are $B_{\mu}^{\dag}|0)$ which correspond to the 1-pair Fermion
states, and the ground state is:
    $$|\Phi_0)=x_1|0)+x_2B_1^{\dag}B_{-1}^{\dag}|0) \, .$$

\section *{III. An Improvement on RPA Based on the Boson description}

\indent
\par
    The excitation elements of RPA are expressed in terms of Fermion
operators and assumed to satisfy the standard Boson commutation rule.
This cause some unphysical state effects which can clearly be seen from a
Boson description. The exact Boson Hamiltonian $H_B$ can be written as
$$H_B=\Gamma_B(\omega_0+\omega_{RPA}\sum\limits_{\mu} A_{\mu}^{\dag}
A_{\mu} )\Gamma_B = \Gamma_B H_A\Gamma_B \, ,$$
where
$$\omega_0=-2\epsilon+\sqrt{(2\epsilon)^2-R^2} \, ,$$
$$ A_{\mu}^{\dag}=X B_{\mu}^{\dag} + Y B_{-\mu}\, ,$$
$$H_A = \omega_0 + \omega_{RPA} \sum\limits_{\mu}A_{\mu}^{\dag}A_{\mu} \, .$$
If we treat the operator $H_A$ as a true Hamiltonian and solve its eigen
equation in the whole Boson state space than we get the RPA results. Such a
kind of eigenstates of $H_A$ of course contain many unphysical components.
    For a further study we now rely on the Boson description to introduce
the RPA in a new point of view. Namely, we regard the following
 as the Boson Hamiltonian of
RPA :
$$H^{RPA}_B=\Gamma_B(\omega_0+\omega_{RPA}\sum\limits_{\mu} B_{\mu}^{\dag}
B_{\mu} )\Gamma_B \, .$$
Next, we will explain the RPA as the zeroth order approximation of our exact
Boson description. For this purpose we make the transformation :
$$\left \{
\begin{array}{l}
{\tilde B}_{\mu}^{\dag}=\beta B_{\mu}^{\dag} +\alpha B_{-\mu}\  \
(\beta^2-\alpha^2=1)\, , \\
B_{\mu}^{\dag}=\beta {\tilde B}_{\mu}^{\dag} -\alpha {\tilde B}_{-\mu}
\end{array} \right. \, ,$$
and write $H_B$ as
$$H_B=\Gamma_B\{E_0^{\prime}+
\omega^{\prime} \sum\limits_{\mu}{\tilde B}_{\mu}^{\dag} {\tilde B}_{\mu}
+R^{\prime}({\tilde B}_1^{\dag}{\tilde B}_{-1}^{\dag} +{\tilde B}_{-1}{\tilde B}_1 )
\}\Gamma_B \, ,$$
with
\begin{eqnarray}
E_0^{\prime} = 4 \epsilon \alpha^2 - 2 \alpha \beta R \, , \nonumber \\
\omega^{\prime}=2\epsilon(\alpha^2+\beta^2)- 2\alpha \beta R \, ,\nonumber \\
R^{\prime} = (\alpha^2 + \beta^2) R - 4 \alpha \beta \epsilon \, .\nonumber
\end{eqnarray}
Introducing $S$ and $\gamma$
$$   S=e^{-\gamma (B_1^{\dag}B_{-1}^{\dag} -B_{-1}B_1)} \, ,$$
$$   e^{\gamma} = \alpha + \beta   \, ,  $$
 $$  e^{- \gamma}  = \beta - \alpha  \, ,   $$
one has
$${\tilde B}_{\mu}^{\dag}=SB_{\mu}^{\dag}S^{\dag} \, , $$
$$
H_B=\Gamma_B S \{E_0^{\prime}+
\omega^{\prime} \sum\limits_{\mu} B_{\mu}^{\dag} B_{\mu}
+R^{\prime}(B_1^{\dag}B_{-1}^{\dag} + B_{-1}B_1 )
\} S^{\dag}\Gamma_B \, .
$$
\par
    If $\alpha $ and $\beta $ are set to be $X$ and $Y$ respectively then
one has
$$H_B=\Gamma_B S_{RPA}(\omega_0+\omega_{RPA}\sum\limits_{\mu} B_{\mu}^{\dag}
B_{\mu} )S_{RPA}^{\dag}\Gamma_B \, .$$
In other word, the RPA Hamiltonan $H^{RPA}_B$ can be found from the expression
for $H_B$ by replacing the operator $S$ with 1 and then determining
$\alpha $ and $\beta $ with the help of the condition $R^{\prime} = 0$. Our
method to improve the RPA is to expand $S$ in powers of $\gamma$ and keep
only a few terms to get an approximate Hamiltonian:
$$H_B \approx \Gamma_B  \{\overline{E_0}+
\overline{\omega} \sum\limits_{\mu} B_{\mu}^{\dag} B_{\mu}
+\overline{R}(B_1^{\dag}B_{-1}^{\dag} + B_{-1}B_1 )
\} \Gamma_B \, .$$
Then the parameters are determined by the condition $ \overline{R} = 0, $ and
the following formulae:
\begin{eqnarray}
\overline{E_0} = E_0^{\prime}+ 2\gamma R^{\prime}+ 2 \gamma^2 \omega^{\prime}
+ \cdots \, , \nonumber \\
\overline{\omega} = \omega^{\prime} + 2\gamma R^{\prime}
+ 2 \gamma^2 \omega^{\prime} + \cdots \, , \nonumber \\
\overline{R} = R^{\prime}+ 2\gamma \omega^{\prime} + 2 \gamma^2 R^{\prime}
+ \cdots \, . \nonumber
\end{eqnarray}
We expect that our simple scheme (MRPA) will be useful in improving
the behaviour of the PRA, especially in the region where the RPA begins to
break down. Numerical results have confirmed our expectation.
We present here results to second order of $\gamma$. We have also
calculated the excitation energy to the fourth order and found it is a little
better. This shows our method is stable and self-consistent.
\par
    The figure is the one Boson excitation energy versus
$\frac{2\epsilon}{R}$. The energy unit is $R$. In order to show more clearly
the improvement  when $\frac{2\epsilon}{R}\rightarrow 1$, the abscissa is
assumed to stand for $\frac{2\epsilon}{R}-1 $ and is scaled logrithmically.
When $\frac{2\epsilon}{R} \rightarrow \infty$, the results of the RPA and our
method are close to the exact solution.
When $\frac{2\epsilon}{R} =1$, both
the RPA and our method give zero excitation energy. In the region where
$\frac{2\epsilon}{R} $ is larger than $1$ but not too large, the one Boson
excitation energy of RPA deviates from the exact solution significantly
and approaches zero rapidly while our method gives a much better description.
The trend our method approaches zero is very slow which is an important
improvement.
\par
\ \par
    This work was supported in part by the National Natural Science
Foundation of China and by Doctoral Programm Foundation of State
Education Commission.

\end{document}